\documentclass[acmsmall]{acmart}
\usepackage{CJKutf8}
\usepackage[utf8]{inputenc}
\usepackage[T1]{fontenc}
\usepackage{textcomp}
\usepackage{multirow}
\usepackage{multicol}
\usepackage{threeparttable} 

\AtBeginDocument{%
  \providecommand\BibTeX{{%
    \normalfont B\kern-0.5em{\scshape i\kern-0.25em b}\kern-0.8em\TeX}}}

\setcopyright{acmcopyright}
\copyrightyear{2022}
\acmYear{2022}
\acmDOI{10.1145/XXXXXXX.XXXXXXX}

\acmJournal{PACMHCI}
\acmVolume{1}
\acmNumber{1}
\acmArticle{1}
\acmMonth{5}

\begin{document}

\title[Understanding How Help-Seeking Posts are Overwhelmed during a Natural Disaster]{"Help! Can You Hear Me?": Understanding How Help-Seeking Posts are Overwhelmed on Social Media during a Natural Disaster}

\author{Changyang He}
\authornote{indicates an equal contribution.}
\affiliation{%
  \institution{Department of Computer Science and Engineering, Hong Kong University of Science and Technology}
  \city{Hong Kong SAR}
  \country{China}
}
\email{cheai@cse.ust.hk}

\author{Yue Deng}
\authornotemark[1]
\affiliation{%
  \institution{Department of Computer Science and Engineering, Hong Kong University of Science and Technology}
  \city{Hong Kong SAR}
  \country{China}
}
\email{ydengbi@connect.ust.hk}

\author{Wenjie Yang}
\affiliation{%
  \institution{Department of Computer Science and Engineering, Hong Kong University of Science and Technology}
  \city{Hong Kong SAR}
  \country{China}
}
\email{wyangbc@connect.ust.hk}

\author{Bo Li}
\affiliation{%
  \institution{Department of Computer Science and Engineering, Hong Kong University of Science and Technology}
  \city{Hong Kong SAR}
  \country{China}
}
\email{bli@cse.ust.hk}

\renewcommand{\shortauthors}{C. He et al.}

\begin{abstract}
Posting help-seeking requests on social media has been broadly adopted by victims during natural disasters to look for urgent rescue and supplies. The help-seeking requests need to get sufficient public attention and be promptly routed to the intended target(s) for timely responses. However, the huge volume and diverse types of crisis-related posts on social media might limit help-seeking requests to receive adequate engagement and lead to their overwhelm. To understand this problem, this work proposes a mixed-methods approach to figure out the overwhelm situation of help-seeking requests, and individuals' and online communities' strategies to cope. We focused on the 2021 Henan Floods in China and collected 141,674 help-seeking posts with the keyword "Henan Rainstorm Mutual Aid" on a popular Chinese social media platform Weibo. The findings indicate that help-seeking posts confront critical challenges of both external overwhelm (i.e., an enormous number of non-help-seeking posts with the help-seeking-related keyword distracting public attention) and internal overwhelm (i.e., attention inequality with 5\% help-seeking posts receiving more than 95\% likes, comments, and shares). We discover linguistic and non-linguistic help-seeking strategies that could help to prevent the overwhelm, such as including contact information, disclosing situational vulnerabilities, using subjective narratives, and structuring help-seeking posts to a normalized syntax. We also illustrate how community members spontaneously work to prevent the overwhelm with their collective wisdom (e.g., norm development through discussion) and collaborative work (e.g., cross-community support). We reflect on how the findings enrich the literature in crisis informatics and raise design implications that facilitate effective help-seeking on social media during natural disasters.
\end{abstract}




\maketitle

\section{INTRODUCTION}

Social media platforms have become crucial information hubs that millions gather during natural disasters. When social media platforms afford crowdsourced information creation and dissemination, people spontaneously spread situational danger warnings and self-rescue knowledge~\cite{qu2009online, qu2011microblogging, olteanu2015expect}, collectively make sense of the situation~\cite{dailey2015s, starbird2013delivering, leavitt2014upvoting,kou2017conspiracy}, collaboratively debunk crisis-related misinformation~\cite{flores2021fighting, hunt2020misinformation}, and mutually provide assistance to facilitate post-disaster recovery~\cite{starbird2011voluntweeters}. Researchers have given increasing attention on how to support effective and efficient crisis communication in HCI and CSCW community~\cite{chen2021exploring,li2021hello,he2021beyond}.

Among various crisis communication needs on social media, \textit{seeking help}, through which victims post urgent help-seeking requests to ask for rescue, supplies or critical information, has become a topic of concern~\cite{han2019using, nishikawa2018time, song2019toward, qu2011microblogging, qu2009online, li2019using, cheng2020explaining, yang2022save}. Compared to directly seeking help from specific agencies (e.g., calling emergency numbers like 911~\cite{conzelmann2005using}), help-seeking posts on social media are broadcast to the public. This approach makes it possible to convey help-seeking requests to broad targets including government emergency offices~\cite{li2021hello, juswil2020government}, non-governmental organizations~\cite{williams2018social} and civil volunteers~\cite{starbird2011voluntweeters} in disasters, who can not be reached through direct help-seeking without knowing their contact information. 

The time sensitivity of post-disaster rescue raises a high demand for help-seeking posts to get sufficient public attention and be promptly routed to the intended target(s)~\cite{keim2011emergent}. However, crisis communication on social media is characterized by \textit{a large volume} and \textit{diversity} of posts~\cite{austin2018social}. For example, Qu et al. revealed that posts that requested help were mixed with many other types of posts that expressed personal feelings, updated situations or raised suggestions~\cite{qu2011microblogging}. Such characteristics of crisis communication might lead to the critical challenge of \textbf{help-seeking overwhelm}, when help-seeking requests fail to receive sufficient public attention and responses and finally get submerged. Understanding this challenge and proposing corresponding strategies would be valuable to facilitate effective help-seeking through social media.

Consequently, this paper investigates how help-seeking posts are overwhelmed during a natural disaster from three perspectives. First, we look into the overwhelm situation of help-seeking posts, aiming to comprehensively uncover the nature of this challenge. Second, when how to express the help-seeking requests substantially influences the transmission of the posts~\cite{luo2020triggers,lifang2020effect,li2021influence}, unearthing the linguistic and non-linguistic strategies\footnote{In this study, we define "linguistic strategies" as help-seeking strategies related to semantics (e.g., detailed description and emotional expression), and use "non-linguistic strategies" to represent the (typically technogized) strategical communication that are less relevant to semantics, such as mentioning others and embedding videos~\cite{veszelszki2017linguistic}.} of individuals in seeking help is of great significance. Third, as an indispensable component of crisis communication on social media~\cite{shklovski2008finding, starbird2013delivering, li2019using, starbird2011voluntweeters}, how online communities respond to the help-seeking overwhelm challenge is also critical yet underexplored. Hence, we propose the following research questions in this work:

\begin{itemize}

  \item \textbf{RQ1}: What is the overwhelm situation of help-seeking posts on social media during a natural disaster?
  
  \item \textbf{RQ2}: What are the strategies that individuals take to prevent help-seeking posts from being overwhelmed, and what are their effects?

  \item \textbf{RQ3}: What are the strategies that online communities use to prevent help-seeking posts from being overwhelmed?
  
\end{itemize}

To understand these research questions, we collected 141,674 posts from a popular Chinese social media platform Weibo that contained the ``Henan Rainstorm Mutual Aid'' keyword during the 2021 Henan Floods in China, and performed a mixed-methods study integrating machine-learning-assisted statistical analysis, regression analysis and qualitative content analysis. We find that help-seeking posts are not only overwhelmed by considerable other categories of posts that also contain the help-seeking-related keyword (we define it as \textit{external overwhelm}), but also face the challenge of attention inequality, with less than 5\% posts attracting more than 95\% public attention (we define it as \textit{internal overwhelm}). We identify a set of linguistic and non-linguistic user-developed strategies for help-seeking. Most of them succeed to promote public engagement (e.g., disclosing danger and vulnerabilities, using subjective narratives and structuring the post to a normalized syntax), yet some fail (e.g., tagging words signifying authenticity in the post). Finally, we reveal the community's spontaneous effort to prevent the overwhelm, which involves both collective wisdom (e.g., norm development through discussion) and collaborative work (e.g., norm enforcement and cross-community support). The findings shed light on design implications to facilitate effective and efficient help-seeking behaviors online during natural disasters.

This work thus enriches the crisis informatics venue in CHI and CSCW community mainly by: (1) revealing the external and internal overwhelm challenges of help-seeking posts during natural disasters; (2) developing a comprehensive taxonomy of linguistic and non-linguistic strategies that influence the public engagement of help-seeking posts; (3) unveiling the community's collaborative work that affords the resilience in response to the help-seeking overwhelm. Under massive and miscellaneous posts on social media-based crisis communication, this work provides insights into future directions to promote the effectiveness of online help-seeking from the perspectives of social-media contexts, individuals and online communities.

\section{RELATED WORK}

\subsection{Crisis Communication on Social Media During Natural Disasters}

Natural disasters are detrimental events resulting from natural processes of the Earth such as hurricanes, floods, firestorms and earthquakes~\cite{alexander2018natural}. Different from public health crises with higher uncertainty~\cite{gui2018multidimensional}, natural disasters typically require straightforward yet more immediate social responses. Today, when social media sites like Facebook and Twitter have become significant information hubs, how crisis communication is performed in social media during natural disasters has attracted great research attention in HCI and CSCW. A large volume of prior work has focused on how the public leveraged social media to curate, spread, and seek critical information in response to natural disasters (e.g., ~\cite{li2021hello, shklovski2008finding, qu2009online, qu2011microblogging, vieweg2010microblogging, olteanu2015expect, madianou2015digital}). A salient feature of crisis communication on social media is the crowdsourced way of information creation and dissemination~\cite{zade2018situational, qu2009online, qu2011microblogging}, which promotes the spread of targeted, timely and situational crisis-related information~\cite{sadri2018crisis}. Also, compared to mass media that largely relies on one-way communication from official response agencies to individuals, social media facilitates bi-directional communication between organizations and the public~\cite{ellison2014cultivating, li2021hello}.

Under the unique characteristics of social media, more nuanced patterns of crisis communication were discovered by crisis informatics researchers. One line of work focused on collective sensemaking ~\cite{leavitt2014upvoting, leavitt2017role, starbird2013delivering, dailey2015s, kou2017conspiracy}. Specifically, when stakeholders naturally converge on the crisis-related social media discussions~\cite{starbird2011voluntweeters, palen2007citizen}, individuals work together to analyze and understand issues and finally increase situational awareness~\cite{dailey2015s, kou2017conspiracy}. Another strand of work focused on deficiencies of social media in crisis communication. For example, some researchers raised the concern about the difficulty in discerning valuable situational information when there was massive user-generated data with uncertain quality~\cite{rudra2015extracting,eriksson2018lessons, rudra2018extracting}, and some work uncovered the prevalence of misinformation and how they negatively influenced stakeholders risk perceptions~\cite{hunt2020monitoring,rajdev2015fake,hunt2020misinformation}.

In general, social media affords miscellaneous crisis communication needs, yet how to improve communication efficiency and efficacy remains a significant challenge for researchers. This study contributes to the venue of crisis communication by comprehensively investigating a critical yet less explored crisis communication type, i.e., help-seeking. We show how help-seeking posts face the challenge of overwhelm, and how individuals and communities develop strategies in response to it.

\subsection{Help-Seeking on Social Media}

Social media has been widely used as a channel for help-seeking when people confront dilemmas (e.g., health problems~\cite{pan2017you, chung2014social}, mental disorders~\cite{pretorius2020searching}, and public health emergencies~\cite{luo2020triggers, han2020using}). During natural disasters, a plethora of research also discovered the extensive use of social media in help-seeking, through which victims urgently published help-seeking posts to look for rescue, supplies, or critical information~\cite{han2019using, nishikawa2018time, song2019toward, qu2011microblogging, qu2009online, li2019using, cheng2020explaining}. Different from other online help-seeking scenarios, help-seeking posts during disasters typically needed to be promptly routed to the intended target(s) such as rescue teams for timely offline responses~\cite{keim2011emergent}. As such, special strategies were developed to increase the attention of help-seeking requests. For instance, the effort of Tweak the Tweet, using a standardized syntax to format the help-seeking posts, emerged as a measure to ease the emergency-related information extraction~\cite{starbird2012promoting}. Another widely-adopted method is including particular hashtags for help-seeking (e.g., "\# HarveySOS" during Harvey Flooding~\cite{yang2017harvey}). Unfortunately, existing communication approaches for help-seeking still face obstacles. For example, Nishikawa et al. showed that most original tweets using the hashtag "\# Rescue" were not for rescue requests during Northern Kyushu floods~\cite{nishikawa2018time}. Song et al. argued that it was not suitable to confirm whether a tweet was a valid rescue request just based on the hashtag, and proposed an automatic detection approach to identify rescue-request tweets~\cite{song2019toward}. These works warn of the inadequate effectiveness of help-seeking online under varied types of posts for crisis communication. 

Some researchers have investigated influencing mechanisms of the spread and response of help-seeking. Content characteristics and creator characteristics were two categories of features that most researchers focused on~\cite{li2021influence, yang2018retweet, wang2019makes}. Relevant features such as content type~\cite{li2021influence, li2021hello}, emotional type and proximity~\cite{luo2020triggers, li2021influence}, depth of self-disclosure~\cite{pan2018you, wingate2020influence}, and social capital of help seekers~\cite{li2021influence} were largely explored on how they influenced the popularity and effectiveness of help-seeking posts. In particular, Luo et al. argued that content factors should be paid more attention to, as most help seekers online were ordinaries with limited followers when requesting rescue~\cite{luo2020triggers}. They investigated how completeness, proximity, support typology, disease severity and emotion may influence the re-transmission of help-seeking microblogs during the COVID-19 pandemic~\cite{luo2020triggers}.

Nevertheless, most prior works on influencing mechanisms of help-seeking were theory-driven, e.g., how negativity bias theory~\cite{rozin2001negativity} was applied in the help-seeking scenario~\cite{lifang2020effect}. What exact user-developed strategies are used to prevent the overwhelm, and whether the strategies work, are still largely under-explored. This work builds a comprehensive taxonomy of linguistic and non-linguistic strategies used by help-seekers and understands how they influence public engagement (RQ2). More importantly, when prior work typically centered on help-seeking as an individual effort (e.g., the influence of content and creator characteristics), how the less-visible work of online communities helps help-seeking posts from being ignored is less investigated. This work also aims to fill this gap (RQ3).

\subsection{Online Communities' Work during Crisis}

When people increasingly seek support and information on social media during disasters, a burgeoning body of research look at how online communities, gathering crowd's effort and wisdom, in turn contribute to crisis communication. One strand of work focused on how online communities helped to speed up the transmission and promote the visibility of critical crisis information (e.g.,~\cite{chen2021exploring, starbird2011voluntweeters, leavitt2017role, st2012trial, park2017framework, kaufhold2016self}). Some special community roles such as digital volunteers and moderators play a crucial role in the process. For example, Starbird and Palen revealed the spontaneous organization and collaboration of digital volunteers in translating, verifying and routing information online in the 2010 Haiti earthquake~\cite{starbird2011voluntweeters}. Leavitt and Robinson uncovered how the gatekeeping mechanism influenced information visibility during crises in Reddit, e.g., how community moderators might consider removing overlapping content to lift the visibility of more crucial information~\cite{leavitt2017role}. 

Another line of work centered on how online communities helped to reduce misinformation and misbehavior during crises (e.g., ~\cite{arif2017closer, he2021beyond, caulfield2020does, zeng2017social, mirbabaie2020breaking}). The wisdom of the crowd is substantial to facilitate such debunking activities~\cite{tanaka2013toward, yang2021know}. For example, Arif et al. examined the ability of self-correction of online communities confronting online rumors during crises, where community members intentionally corrected themselves, the information space, or other users~\cite{arif2017closer}. Qu et al. uncovered that users shaped community norms and requested for moderation to regulate others' misbehavior in the 2008 Sichuan Earthquake in the Chinese online forum \textit{Tianya}~\cite{qu2009online}. He et al. demonstrated that video viewers collectively leveraged synchronous commenting features to correct misinformation in COVID-19-related videos~\cite{he2021beyond}.

Recent evidence indicated that online communities also played an important role to prevent help-seeking posts from getting overwhelmed by massive user-generated data during crises~\cite{ZhengzhouNews}. Nonetheless, which specific strategies are developed, and how they are put into practice to prevent the overwhelm, are still under-investigated. This study performed qualitative content analysis to reveal online communities' collective wisdom and collaborative work to prevent the help-seeking overwhelm in depth.

\section{METHOD}

\begin{figure}
	\centering
		\includegraphics[scale=.48]{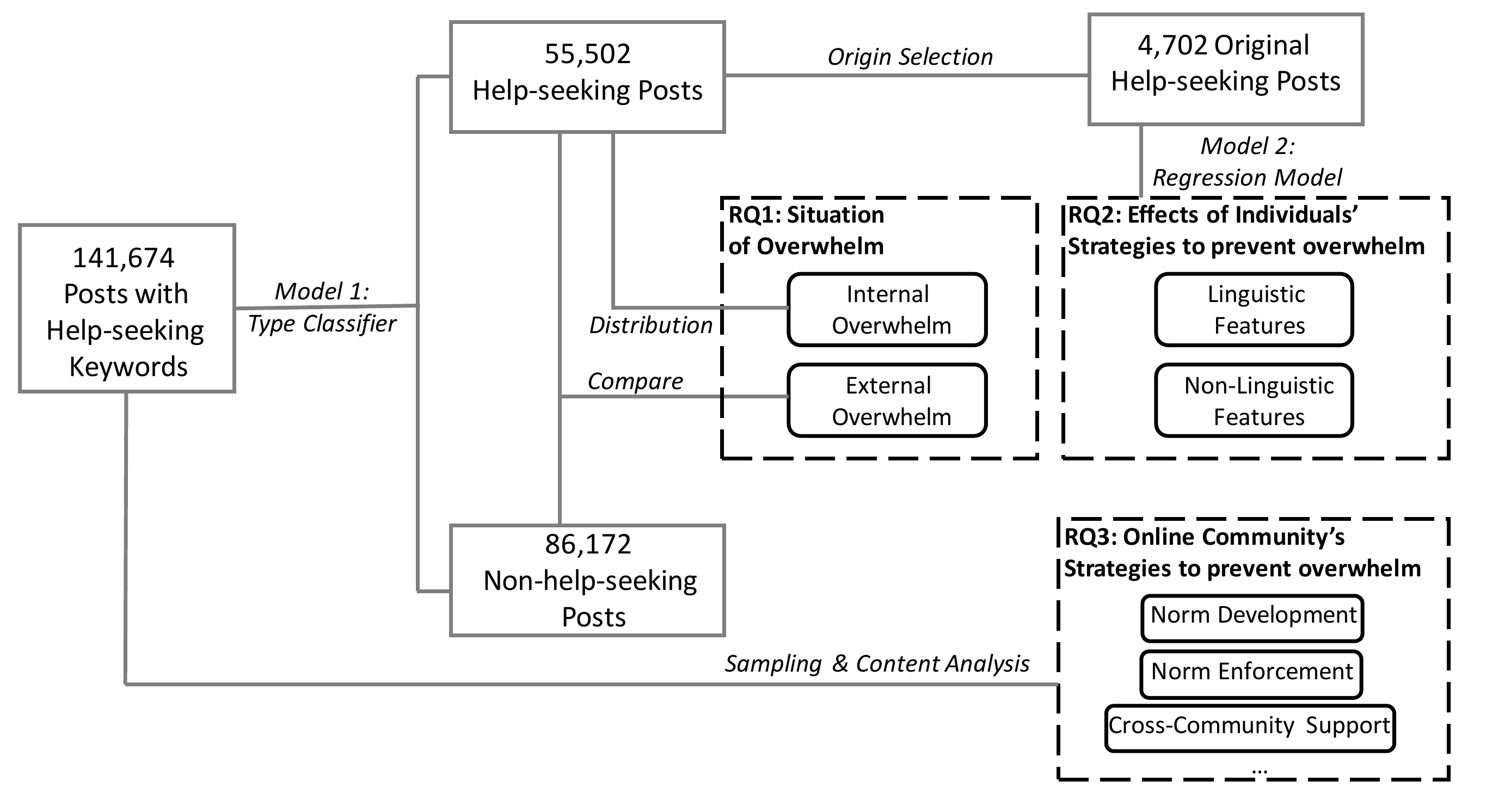}
	\caption{The overall analytical flow to understand how help-seeking posts are overwhelmed during the 2021 Henan Floods}
	\label{FIG: flow-chart}
\end{figure}

We adopted a mixed-methods study to systematically understand how help-seeking posts were overwhelmed during a natural disaster. We first described the study event (Section \ref{event}) and data selection and collection (Section \ref{collection}) to contextualize the investigation. Then, we reported how we leveraged machine-learning-assisted statistical analysis to figure out the overwhelm situation (RQ1, Section \ref{RQ1-METHOD}), applied regression analysis to unveil the effects of individuals' linguistic and non-linguistic strategies to prevent overwhelm (RQ2, Section \ref{RQ2-METHOD}), and used qualitative content analysis to understand the community's overwhelm-prevention strategies (RQ3, Section \ref{RQ3-METHOD}). The overall analytical flow is shown in Figure \ref{FIG: flow-chart}.

\subsection{Study Event: the 2021 Henan Floods}\label{event}

The 2021 Henan Floods occurred in Henan province, China throughout July 2021~\cite{HenanFloods}. A prolonged period of heavy rainfall was the cause of the event, where the peak of rainfall reached 201.9 millimeters (7.95 inches) within an hour. The floods caused widespread damage. As of 2 August 2021, about 14.5 million people were affected by the floods across 150 counties, cities and districts including Zhengzhou, Xinxiang and Weihui, and 815,000 residents were evacuated\footnote{https://www.henandaily.cn/content/2021/0802/312251.html}. Provincial authorities reported 302 deaths, and over 50 were missing\footnote{https://www.163.com/news/article/GGDLBBVE0001899O.html}. 

Weibo, the largest Chinese microblogging website, has about 10 years' history and grew as one of the dominant sites for crisis communication in China~\cite{yang2021know, chen2021exploring, qu2011microblogging}. Similar to Twitter, users could post content in text, picture or video, and others could interact through like, comment or share (retweet) interfaces. Posts with more likes, comments and shares have a higher popularity score on Weibo\footnote{https://www.weibo.com/ttarticle/p/show?id=2309404007731978739654}. As such, they are intuitively more visible under the "TOP" recommendation mode similar to Twitter. These actively-interacted posts also have a higher chance to appear on the default "real-time microblog" of a specific topic or search, when posts are "\textit{ranked by the latest share or like timestamp}" as indicated in the interface. Weibo played a crucial role in help-seeking information posting and transmission During the 2021 Henan Floods. As of November 2021, the posts with hashtag "\textit{\#Henan Rainstorm Mutual Aid}" had reached 24.6 million (including retweets and comments), receiving a total of 17 billion read\footnote{https://s.weibo.com/weibo?q=\%23\%E6\%B2\%B3\%E5\%8D\%97\%E6\%9A\%B4\%E9\%9B\%A8\%E4\%BA\%92\%E5\%8A\%A9\%23}. 

\subsection{Data Selection and Collection}\label{collection}

Before collecting the data, we went through posts regarding the 2021 Henan Floods on the Weibo platform to get a preliminary understanding of users' online help-seeking behavior in this context. We noticed that users developed diverse strategies to identify the help-seeking purpose of their posts: (1) use the hashtag "\textit{\#Henan Rainstorm Mutual Aid\#}"; (2) contain the Super Topic "\textit{Henan Rainstorm Mutual Aid}", through which interface users could follow specific topics and form long-lasting communities~\cite{SuperTopic, chen2021exploring}; (3) only include the plain text "\textit{Henan Rainstorm Mutual Aid}". Consequently, to make the data comprehensive, we chose keyword-based collection ("\textit{Henan Rainstorm Mutual Aid}") to identify all help-seeking posts instead of the hashtag- or super-topic-based collection approach.

We performed the keyword-based data crawling assisted by the WeiboSuperSpider tool~\cite{SuperTopicCrawling}. The date ranged from July 17, 2021, the beginning of the Henan heavy rain~\cite{HenanFloods}, to August 8, 2021, when the provincial emergency response of flood control reduced from level I to level III~\cite{HenanFloodsBaidu}. We collected the data on September 28, 2021, a sufficiently long time after the end of the date range. In this way, the social media engagement indexes (i.e., likes, comments and shares)~\cite{chen2020unpacking,brubaker2018let,denktacs2020stakeholder} became steady for a relatively fair comparison between microblogs posted in different periods. However, it inevitably led to the loss of part of the data due to post deletion, especially when "\textit{deleting posts after getting rescued}" was developed as a norm by the community to promote the visibility of less-noticed help-seeking posts (see Section \ref{communityWork}).

In total, we retrieved 141,674 posts with the keyword "\textit{Henan Rainstorm Mutual Aid}" contributed by 72,075 distinct users, including 13,591 original posts from 9,996 users. 

\subsection{RQ1: Understanding the Overwhelm Situation}\label{RQ1-METHOD}

The exploratory observation of help-seeking posts before data collection empirically indicated two types of overwhelm: (1) \textbf{external overwhelm}, the challenge of being overwhelmed by a large volume of \textit{non-help-seeking} posts that also included the help-seeking-related keyword (i.e., "\textit{Henan Rainstorm Mutual Aid}"); (2) \textbf{internal overwhelm}, the challenge of being overwhelmed by numerous other help-seeking posts. We designed specialized strategies to investigate the two overwhelm types.

\subsubsection{External Overwhelm}\label{externalM}

To quantify different types of posts with the "\textit{Henan Rainstorm Mutual Aid}" keyword, we manually generated a codebook through corpus annotation, and then built a multi-class text-classifier to assign the codes to the whole corpus.

We first tried to figure out the types of posts in our dataset and obtain a training dataset. We used inductive thematic analysis to annotate the data~\cite{fereday2006demonstrating}, letting codes naturally emerge. Specifically, two authors first independently read 200 samples and generated initial codes (e.g., "\textit{rescue seeking}" and "\textit{situational risk reminders}") that directly reflected the data. Through discussion and comparison, they merged the initial codes and concluded the final codebook with high-level types (e.g., "\textit{seeking help}" and "\textit{offering help}"). The agreement ratio (93.5\%) and Cohen's Kappa ($\kappa = 0.90$) indicated substantial agreement between the two coders~\cite{mchugh2012interrater}. Disagreement was resolved through another round of discussion. Finally, two annotators separately coded another 400 posts each, reaching a training dataset with a total of 1000 type-assigned samples. The codebook is shown in Table \ref{tab:taxonomy}.

After building the training dataset, we developed the type classifier as follows:

\begin{itemize}
  \item \textbf{Preprocessing}: We used the Jieba word segmentation tool to segment the Chinese text~\cite{jieba}, and removed stopwords based on the HIT Chinese stopwords table~\cite{stopword}. Punctuations, numbers and URLs were also excluded.
  \item \textbf{Word Embedding}: We trained the Word2Vec word embedding model~\cite{mikolov2013distributed} on the whole dataset, representing each word as a 300-dimension vector.
  \item \textbf{Model Selection}: We tried both traditional machine learning models (e.g., SVM and xgboost) through averaging word vectors to vectorize posts, and deep learning models (GRU and LSTM). We finally chose LSTM~\cite{hochreiter1997long} for its best performance. Specifically, under 10-fold cross-validation, the micro F1 score for the 4-class classification achieved 79.6\%, which was substantially good to generalize the human annotations to the whole dataset. In the training, we used Adam as the optimizer, and added a dropout layer (rate=0.2) to prevent overfitting when our labeled data was limited~\cite{gal2016theoretically}.
\end{itemize}

We applied the classifier to assign codes to all posts. To explore the external overwhelm by \textit{non-help-seeking} posts, we compared the volume and engagement indexes (likes, comments and shares) of \textit{seeking help} posts to other three post types in the dataset.

\begin{table*}
  \small
  \caption{The codebook of post types with the help-seeking-related keyword ``\textit{Henan Rainstorm Mutual Aid}'' during the Henan Floods (the keyword is omitted in the examples)}
  \label{tab:taxonomy}
  \begin{tabular}{p{2cm}p{3.3cm}p{5.8cm}p{1.5cm}}
    \toprule
    Type & Definition & Example & Percentage in the Sample \\
    \midrule
    Seeking Help & Posts requesting help especially rescue and supplies & \textit{[The location] has at least 24 children (the youngest is less than half a year old) and 50 adults. No water, no power, no gas, poor signal. We need water and food, and if possible, getting evacuated. [phone number]} & 33.9\% \\
    
    Offering Help & Posts about offering help to victims and rescue teams & \textit{I am now in Zhengzhou, and plan to go to the disaster areas of Xinxiang and Weihui. I can lead the way for any official or civilian rescue team. I am very familiar with the disaster area! Contact me: [phone number]} & 9.9\% \\
    
    Transmitting Critical Information & Posts sharing critical disaster-related information such as self-rescue knowledge, situational risk reminders and community norms & \textit{Just received the information. Water short-circuiting happened in [location]. Three people died from the electric shock. Remind again: Never go out at random now. The Road is flooded. It is very dangerous!} & 15.6\% \\
    
    Sharing Attitudes and Opinions & Posts sharing personal attitudes and opinions, such as blessing and raising attention & \textit{Wish everyone safe and peaceful!} & 40.6\%\\
    
    \bottomrule
  \end{tabular}
\end{table*}

\subsubsection{Internal Overwhelm}

Based on the type classifier, we selected the pure \textit{seeking help} posts to focus on the internal overwhelm, i.e., how they were overwhelmed by each other, as shown in Figure \ref{FIG: flow-chart}. We investigated the distribution of the user-engagement indexes to figure out the attention inequality~\cite{zhu2016attention} of \textit{seeking help} posts. In particular, we referenced the Pareto principle (80/20 rule)~\cite{dunford2014pareto} and Gini Index~\cite{farris2010gini} to measure the inequality, which were initially adopted in the social and economic area, and recently used to test inequality in social media~\cite{van2016employing, zhu2016attention, indaco2016urban}.

\subsection{RQ2: Investigating Effects of Individuals' Strategies to Prevent Overwhelm}\label{RQ2-METHOD}

In this section, we described how we identified individuals' linguistic and non-linguistic strategies to prevent overwhelm, and leveraged regression analysis to comprehensively figure out their effects.

\subsubsection{Dataset Consideration}

We selected original help-seeking posts as the research target for RQ2, as (1) they were representative to capture individuals' linguistic and non-linguistic strategies, and (2) the origin selection helped to avoid the influence of repetitive content in shared posts to regression analysis. To achieve so, we first filtered out non-help-seeking posts based on the type classifier developed in Section \ref{RQ1-METHOD}, and selected the original ones, as shown in Figure \ref{FIG: flow-chart}. This step yielded 4,702 original help-seeking posts for further analysis.

\subsubsection{Identifying Individuals' Linguistic and Non-linguistic Strategies}\label{strategy_extraction}

Linguistic patterns, such as emotions, proximity and information completeness, directly affected how people perceived and responded to crisis information~\cite{luo2020triggers,li2021hello,li2021influence}. Recently, non-linguistic strategies, such as enhancing media richness (e.g., using images and videos)~\cite{chen2020unpacking} and structuring the post to a fixed format~\cite{starbird2011voluntweeters}, were also discovered in crisis informatics literature. Thus, we performed a grounded-theory-based approach~\cite{ritzer2007blackwell} to identify which linguistic and non-linguistic strategies were adopted by individuals when drafting seeking-help requests to attract public engagement. Note that many strategies were unconsciously used by help-seekers when they described the situations, such as disclosing the vulnerability and severity of disasters to draw attention. Specifically, two authors coded 200 samples to abstract potential linguistic and non-linguistic strategies separately (saturation was reached after coding 80-100 posts). They merged similar codes and resolved the disagreement through discussion, and iterated for several rounds to reach a consensus.

After identifying the strategies, we then focused on their patterns for further automatic extraction, which was a prerequisite step for building the regression model. Based on the nature of each linguistic and non-linguistic pattern, we either chose regular expression matching or existing extraction tools. We iteratively (1) derived patterns from the data, (2) designed corresponding regular expressions or searched for suitable tools, (3) applied regular expression or tools to extract the feature, and (4) manually validated a set of samples to ensure the accuracy of our extraction. Finally, all strategy-extraction approaches reached an accuracy rate above 80\% based on our manual validation on 100 random samples, with most of them higher than 95\%, indicating sufficiently precise feature extraction.

\begin{figure}
	\centering
		\includegraphics[scale=.35]{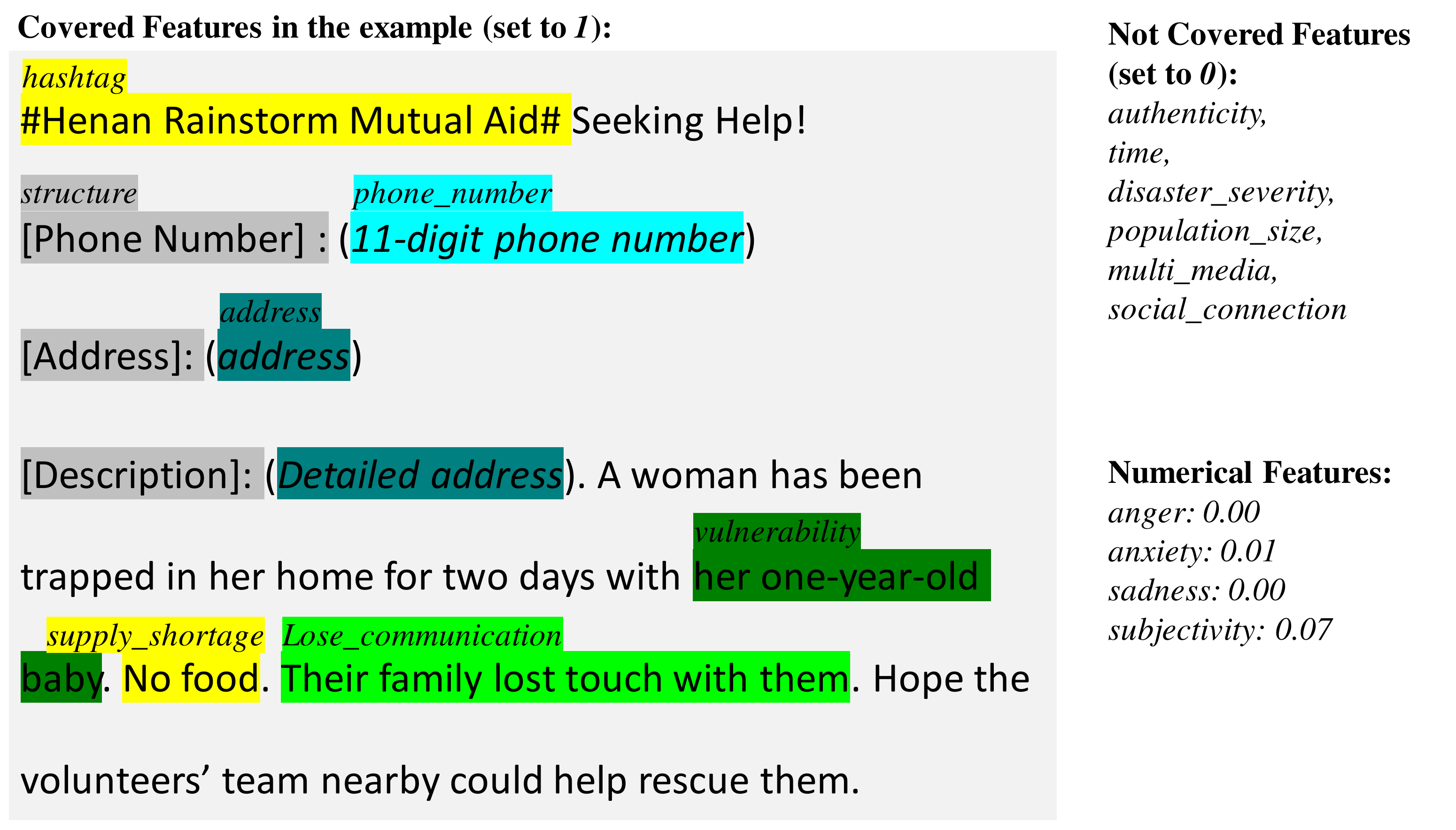}
	\caption{A help-seeking post example that illustrates the feature extraction of linguistic and non-linguistic strategies.}
	\label{FIG: linguistic}
\end{figure}

We provide the description as well as the extraction method for each strategy as follows. A typical example of our feature extraction of linguistic and non-linguistic strategies is shown in Figure \ref{FIG: linguistic}.

\begin{itemize}

  \item \textit{\textbf{Linguistic Features}}
  \begin{itemize}
      
      \item \textbf{Authenticity} (binary): The strategy of explicitly tagging the seeking-help post with keywords signifying authenticity. It was extracted from keyword matching, e.g., "\textit{verified}".
    %   It was a common strategy that emerged from the data when the victim (or the volunteer representing the victim) aimed to emphasize the authenticity.
      
      \item \textbf{Information Detailedness}. Providing adequately detailed information (\emph{phone\_number}, \emph{address} and \emph{time}) to facilitate the rescue.
          \begin{itemize}
          \item \emph{phone\_number} (binary): Whether a post contained the phone number of the contact person. It was extracted by 11-digit-number regex matching.
         \item \emph{address} (binary): Whether a post contained the specific location to perform the rescue. It was identified by cocoNLP\footnote{https://github.com/fighting41love/cocoNLP}, a tool with good performance to match time and addresses in Chinese.
          \item \emph{time} (binary): Whether a post described the exact time of help-seeking. It was extracted by cocoNLP.
          \end{itemize}

      \item \textbf{Danger}. The strategy of disclosing situational danger (\emph{vulnerability}, \emph{supply\_shortage}, \emph{disaster\_severity}, \emph{population\_size} and \emph{lose\_communication}) to draw attention from the public. If not specified, the feature was extracted through regex matching.
          \begin{itemize}
          \item \emph{vulnerability} (binary): Whether a post disclosed vulnerable groups, such as children, seniors, patients and pregnant women. 
          \item \emph{supply\_shortage} (binary): Whether a post indicated supply shortage, including lack of water, electricity, gas and food.
          \item \emph{disaster\_severity} (binary): Whether a post directly described the flood situation (i.e., rising, retreating, the depth of water, etc). It was extracted by keywords (e.g., "rising") or number matching (for water level) supported by Sinan\footnote{https://github.com/yiyujianghu/sinan}.
          \item \emph{population\_size} (numeric): The size of affected population extracted by Sinan (in a logarithmic scale).
          \item \emph{lose\_communication} (binary): Whether a post showed the potential danger of losing communication, e.g., mobile phones out of charge or signal.
          \end{itemize}
          
      \item \textbf{Emotion}. Disclosing negative emotions. The negativity bias theory suggests that negative emotions have a stronger impact on a person's behavior and cognition~\cite{rozin2001negativity}. \emph{Anger}, \emph{anxiety} and \emph{sadness} well captured the negative emotional disclosure in our context, which also echoed prior work~\cite{luo2020triggers,lifang2020effect}. We applied Jingdong Chinese Sentiment API\footnote{https://neuhub.jd.com/ai/api/nlp/sentiment} to generate the emotion score.
          \begin{itemize}
          \item \emph{anger} (numeric): The level of anger in the post. 
          \item \emph{anxiety} (numeric): The level of anxiety in the post. 
          \item \emph{sadness} (numeric): The level of sadness in the post. 
          \end{itemize}
      \item \textbf{Subjectivity} (numeric): Using subjective narratives to ask for help. It was calculated by TextBlob\footnote{https://textblob.readthedocs.io/en/dev/quickstart.html\#sentiment-analysis}, an effective tool of subjectivity inference, after Chinese-English translation.
  \end{itemize}
  
  \item \textit{\textbf{Non-linguistic Features}}
  \begin{itemize}
  
  \item \emph{structure} (binary): Whether structuring the post to a normalized syntax to facilitate information extraction~\cite{starbird2011voluntweeters, starbird2012promoting}. The normalized syntax, proposed by digital volunteers, encouraged help-seekers to structure the details of help-seeking information (e.g., phone number, address, help-seeking time, help-seeker's name and situation description) line by line, and use brackets to enclose detail labels for clarity, as shown in Figure \ref{FIG: linguistic}. It was identified by regex matching (combining bracket and keyword pairing).
  \item \emph{multi\_media} (binary): Whether using images or videos to describe the disaster situation. It was an intrinsic feature in our crawled data.
  \item \emph{hashtag} (binary): Whether using hashtag(s) to identify the help-seeking requests (extracted by hashtag matching).
  \item \emph{social\_connection} (binary): Whether mentioning others (extracted by matching "@").
  \end{itemize}
  
\end{itemize}

\subsubsection{Developing Regression Model}

We developed regression models to understand the effects of the identified linguistic and non-linguistic strategies, i.e., whether they successfully promoted public engagement and prevented the overwhelm. We took likes, comments, shares and their sum as dependent variables to measure the social media engagement~\cite{brubaker2018let,kim2017like,chen2020unpacking}. The distributions of these indexes were over-dispersed (likes: M=26.53, SD=404.43; comments: M=6.87, SD=58.97; shares: M=46.55, SD=709.41), when the conditional variance exceeds the conditional mean. Therefore, we applied the negative binomial regression~\cite{hilbe2011negative}, the appropriate regression model to process over-dispersed count data, to analyse how the strategies influenced public engagement.

We also considered \textit{post type} and \textit{contextual features}, which were proved influential to public engagement during disasters~\cite{luo2020triggers,li2021influence}, as two categories of control variables in the regression models.

\begin{itemize}

    \item \textbf{Type}: Whether the help-seeking post was used to \textit{seek rescue} or \textit{seek supply}. The method to establish the classification criteria was similar to the pattern-extraction approach in Section \ref{strategy_extraction}. Specifically, as we observed that most \textit{seek supply} posts explicitly used "need/badly need (something)" keywords (e.g., "need pumps") to describe what they lacked, while \textit{seek rescue} posts did not (and typically made strong requests such as "asking for help" or "seeking help"), we identified the post type using need-relevant keywords. This classification approach was proved practical when the accuracy reached 81\% on the 100 manual-validation sample. Note that different from public health crises in which people frequently asked for situational information due to the high uncertainty~\cite{gui2017managing, luo2020triggers}, \textit{seeking information} was barely discovered in our data samples.
    
    \item \textbf{Contextual Features}: Number of \emph{follower} and \emph{following} of the help-seeker, and the number of help-seeking posts in the post date (\emph{information\_density}). These features naturally defined the context of crisis communication~\cite{li2021hello}.
\end{itemize}

\subsection{RQ3: Exploring the Online Community's Strategies to Prevent Overwhelm}\label{RQ3-METHOD}

We performed qualitative content analysis to understand the online community's strategies to prevent help-seeking posts from being overwhelmed. Particularly, we focused on both (1) the \textit{strategy practices}, which were perceived in interactions that directly reflected the community' support; and (2) the \textit{strategy development}, which were identified from discussions among community members, especially when some users realized the problem of help-seeking overwhelm and tried to think out community-based support. We focused on the two aspects instead of only strategy practices considering some community's work was invisible, e.g., the coordination of digital volunteers. As such, we sampled the posts for content analysis from the whole dataset instead of the filtered help-seeking posts as shown in Figure \ref{FIG: flow-chart}, when \textit{strategy development} largely came from non-help-seeking posts (e.g., sharing attitudes). Posts with heated discussion in replies, which well captured the community's collaborative work, were more focused. Two authors independently coded 100 random posts and their comments to generate initial codes. Then, they discussed and compared the codes to reach a consensus, and went back to code additional data, iterating for several rounds till no new codes emerged. In total, they coded 220 posts with 1423 replies.

\section{FINDINGS}

In this section, we reported the findings on the overwhelm challenge of help-seeking posts during a natural disaster. In Section \ref{overwhelmSituation}, we uncovered the overwhelm situation of help-seeking posts, showing that the engagement of help-seeking requests was not only affected by enormous non-help-seeking posts which also contained the help-seeking-related keyword, but also threatened by the attention inequality with very few help-seeking posts receiving most responses. In Section \ref{influencingMechanism}, we described how individuals' linguistic and non-linguistic strategies to prevent overwhelm influenced public engagement. In Section \ref{communityWork}, we illustrated the strategies that were spontaneously developed by the online community to prevent the overwhelm supported by its collective wisdom and collaborative work. The findings revealed the rich nuances of the overwhelm challenge of seeking help on social media during a natural disaster, and shed light on potential countermeasures from social-media contextual (RQ1), individual (RQ2) and community (RQ3) perspectives.

\subsection{RQ1: Overwhelm Situation}\label{overwhelmSituation}

\subsubsection{External Overwhelm}\label{ExternalResult}

\begin{figure}
	\centering
		\includegraphics[scale=.47]{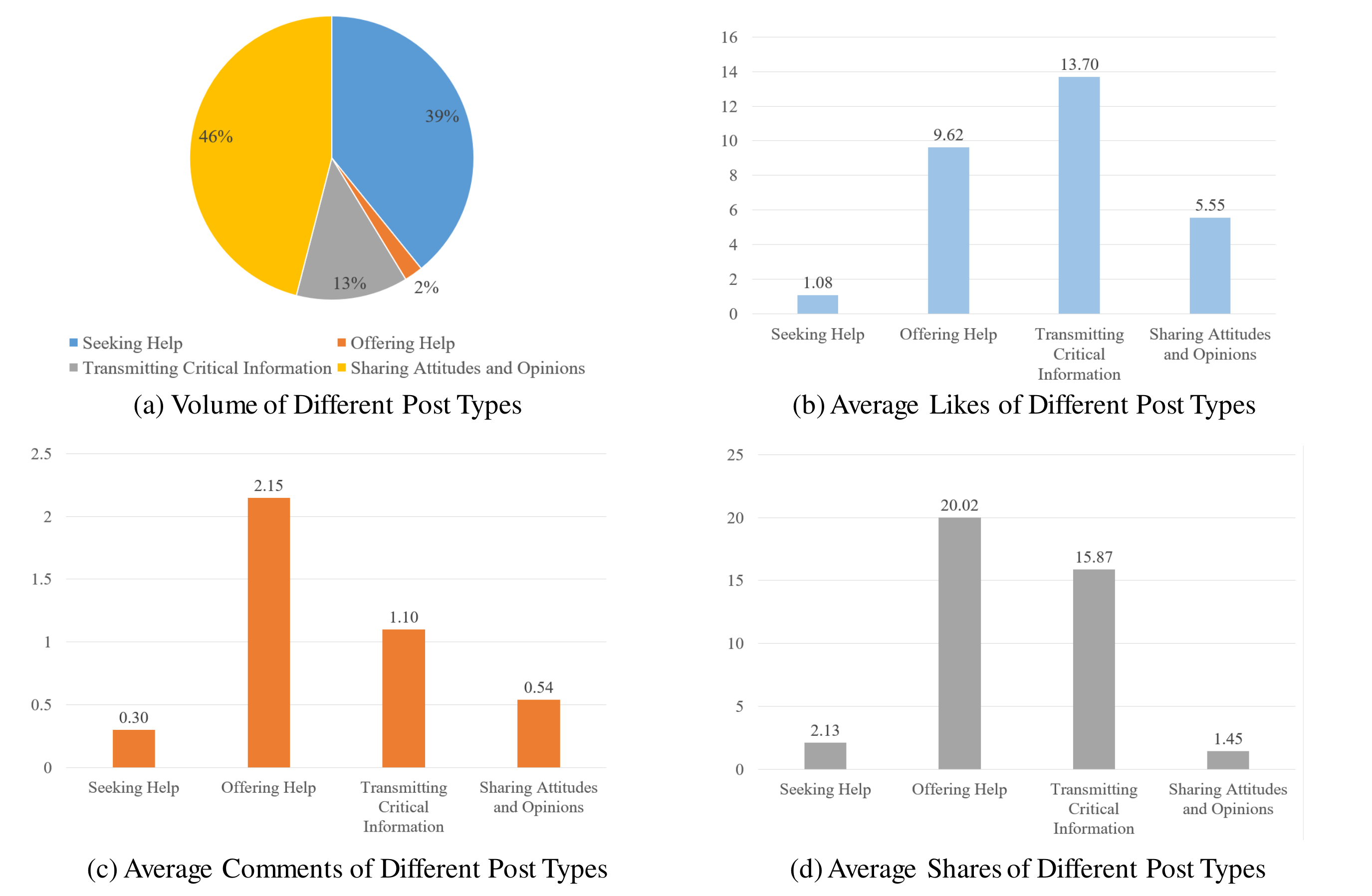}
	\caption{The comparison of different post types in (a) Volume, (b) Likes, (c) Comments, and (d) Shares.}
	\label{FIG: External}
\end{figure}

The comparison between \textit{seeking help} posts and the other three post categories in volume, likes, comments and shares is shown in Figure \ref{FIG: External}. For all engagement indexes (likes, comments and shares), the null hypotheses of the one-way ANOVA test (i.e., all post types had the same means)~\cite{kim2014analysis} were rejected ($p<0.05$), indicating there existed substantial inter-category differences.

The comparison yielded several significant findings: (1) The \textit{seeking help} posts received the fewest likes and comments, and were shared much less than posts \textit{offering help} or \textit{transmitting critical information}, even though the keyword for data collection "\textit{Henan Rainstorm Mutual Aid}" was mainly designed for helping those in need during crises~\cite{ZhengzhouNews}; (2) The \textit{seeking help} posts were also not the dominant category in amount (39.2\%, $N = 55,502$), which was less than posts \textit{sharing attitudes and opinions}; (3) The \textit{offering help} posts, though not large in volume (2.1\%, $N = 3,003$), received the most average comments and shares, exhibiting people's particular engagement on this category of posts; (4) The \textit{transmitting critical information} posts were liked most, showing public's acceptance and gratitude toward such kind of posts. 

Note that the comparison of help-seeking posts with the other three post categories did not hint that the other three post categories were less important. Indeed, posts that offered help (e.g., providing the phone number of the rescue teams) and transmitted critical information (e.g., which areas were dangerous during the crisis) were a crucial component of crisis communication~\cite{qu2011microblogging}. The \textit{sharing attitudes and opinions} posts would also help to attract people's attention from the whole microblogging platform. However, our findings revealed the risk of the overwhelm of individual's online help-seeking by miscellaneous information during the natural disaster, especially when no interface helped to select the pure help-seeking posts.

\subsubsection{Internal Overwhelm}\label{internalResult}

\begin{figure}
	\centering
		\includegraphics[scale=.4]{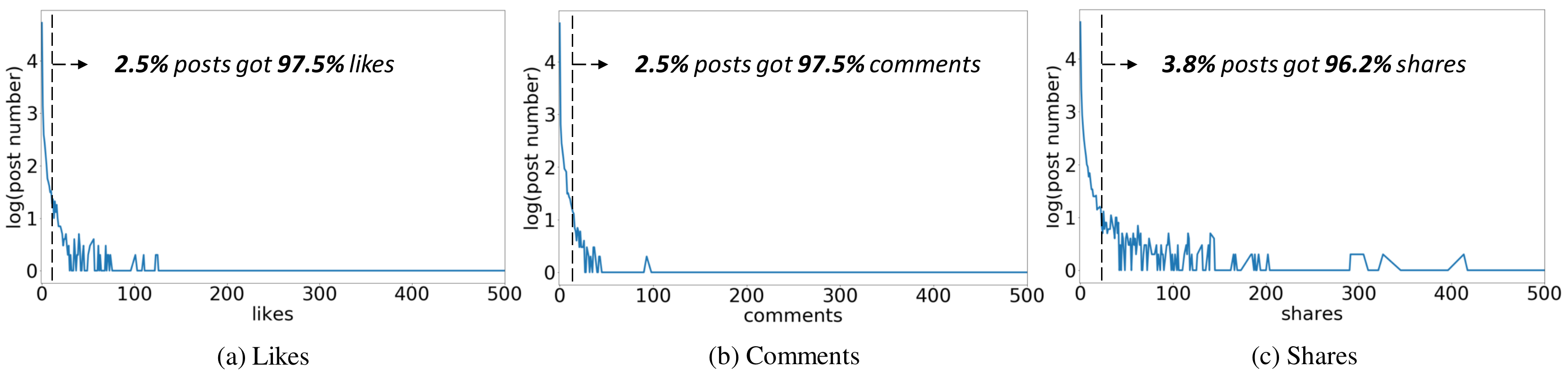}
	\caption{The distribution of \textit{seeking help} posts (after a logarithmic scale of the post volume) in (a) Likes, (b) Comments, and (c) Shares. For all three engagement indexes, less than 5\% posts got more than 95\% attention, indicating extreme imbalance and the high chance of being overwhelmed for individuals' help-seeking.}
	\label{FIG: Internal}
\end{figure}

Figure \ref{FIG: Internal} demonstrated the distribution of pure \textit{seeking help} posts in likes, comments, and shares after a logarithmic scale of the post volume. The results signified extreme imbalance of user attention: 2.5\% posts got 97.5\% likes and comments, and 3.8\% posts got 96.2\% shares, which was more imbalanced than the Pareto principle (80/20 rule)~\cite{dunford2014pareto} and previous findings on general Twitter posts~\cite{zhu2016attention}. In particular, 94.9\%, 96.7\%, and 43.3\% \textit{seeking help} posts got 0 like, comment and share respectively, and 41.8\% among all posts received 0 in all three indexes, indicating the high proportion of help-seeking posts with little or zero attention. The Gini indexes for likes, comments, and shares of help-seeking posts were 0.995, 0.992, and 0.989, which also indicated the extreme engagement inequality.

These findings largely revealed the nature of attention inequality for \textit{seeking help} posts during the natural disaster. Based on that, the following sections \ref{influencingMechanism} and \ref{communityWork} demonstrated individuals' and the online community's effort to prevent overwhelm.

\subsection{RQ2: Effects of Individuals' Linguistic and Non-linguistic Strategies}\label{influencingMechanism}

We implemented negative binomial regression models to explore how \textit{linguistic} and \textit{non-linguistic} strategies affected public engagement of help-seeking posts in the natural disaster. The results of the negative binomial regression are shown in Table \ref{tab: performance_comparison}. 

\begin{table}[htbp]
	\footnotesize
	\centering
	\caption{Results of negative binomial regressions for social media engagement. IRR (Incidence Rate Ratio) indicates the ratio change of the dependent variable when increasing an independent variable by one unit. For a specific feature, if IRRs across all engagement dimensions (likes, comments, shares and their sum) are great than 1, we italicize the feature to denote its strong promotion effect. ***p\textless0.001; **p\textless0.01; *p\textless0.05.}
	\label{tab: performance_comparison}
	\begin{tabular}{p{2cm}p{2.4cm}p{0.5cm}p{1cm}p{0.5cm}p{1cm}p{0.5cm}p{1cm}|p{0.5cm}p{1cm}}
		%\toprule
		\hline
		& & \multicolumn{2}{l}{Likes}&\multicolumn{2}{l}{Comments} & \multicolumn{2}{l}{Shares}&\multicolumn{2}{|l}{Sum}\\
		\cline{3-10}
		& &IRR&Std. Err.&IRR&Std. Err.&IRR&Std. Err.&IRR&Std. Err.\\
		%\midrule
		\hline
		
		\multicolumn{10}{l}{\textbf{Linguistic Strategies}}\\
		\hline
		Authenticity&authenticity&0.91***&0.017&0.94***&0.018&0.96*&0.017&0.92***&0.017\\
		
        \cline{1-2}
		\multirow{3}{1in}{Detailedness} 
		&\textit{phone\_number}&1.16***&0.018&1.14***&0.019&1.33***&0.018&1.31***&0.018\\
		&address&0.96*&0.016&1.07***&0.017&1.03&0.016&0.97&0.016\\
		&\textit{time}&1.07***&0.016&1.04*&0.017&1.18***&0.015&1.07***&0.015\\
		%\hline
		
		\cline{1-2}
		\multirow{5}{1in}{Danger} 
		&vulnerability&0.93***&0.016&1.13***&0.017&1.10***&0.016&1.01&0.016\\
		&supply\_shortage&0.91***&0.019&0.95**&0.019&0.73***&0.018&0.87***&0.018\\
		&\textit{disaster\_severity}&1.10***&0.017&1.12***&0.017&1.18***&0.016&1.12***&0.016\\
		&population\_size&0.95**&0.017&0.95**&0.017&1.14***&0.016&1.04*&0.016\\
		&\textit{lose\_communication}&1.16***&0.018&1.13***&0.018&1.43***&0.017&1.23***&0.017\\
		
		\cline{1-2}
		\multirow{3}{1in}{Emotion} 
		&anger&0.96**&0.017&0.96*&0.018&0.73***&0.018&0.87***&0.017\\
		&anxiety&0.95**&0.017&0.99&0.018&1.05**&0.017&1.05**&0.017\\
		&sadness&0.94**&0.02&1.02&0.019&0.99&0.023&0.96*&0.018\\
		
		\cline{1-2}
		Subjectivity&\textit{subjectivity}&1.13***&0.016&1.06***&0.016&1.09***&0.015&1.13***&0.015\\
		\hline
		
		\multicolumn{10}{l}{\textbf{Non-linguistic Strategies}}\\

		\hline
    	\multirow{4}{1in}{Non-linguistic features}
		&\textit{structure}&1.07***&0.016&1.01&0.017&1.10***&0.016&1.06***&0.016\\
		&\textit{multi\_media}&1.36***&0.016&1.31***&0.016&1.62***&0.015&1.39***&0.015\\
		&\textit{hashtag}&1.51***&0.016&1.04*&0.017&1.35***&0.015&1.39***&0.015\\
		&social\_connection&0.91***&0.016&0.96*&0.017&0.95***&0.015&0.93***&0.015\\
		
		\hline
		\multicolumn{10}{l}{\textbf{Control Variables}}\\

        \hline
		Type&seeking\_supply (reference group: seeking\_rescue)&0.78***&0.018&0.84***&0.019&0.85*&0.017&0.79***&0.017\\

		\cline{1-2}
		\multirow{3}{1in}{Contextual features} 
		&\textit{follower}&3.00***&0.015&1.58***&0.015&2.59***&0.015&2.57***&0.015\\
		&\textit{following}&1.22***&0.015&1.22***&0.015&1.39***&0.015&1.29***&0.015\\
		&information\_density&0.66***&0.016&0.75***&0.017&0.85***&0.015&0.75***&0.015\\
		\hline
		%\bottomrule
	\end{tabular}
	\label{proportion}
\end{table}

\subsubsection{Linguistic Strategies}

The investigation on the five linguistic strategies (\textit{tagging authenticity, providing detailed rescue-relevant information, disclosing danger, expressing negative emotions} and \textit{using subjective narratives}) indicated their substantial influences on how people responded to the help-seeking requests. Specifically, we concluded the following findings: (1) Manually tagging the \textit{authenticity} in the post had no promotion of public attention; (2) How users perceived and responded to help-seeking posts were contingent upon the \textit{detailedness} of the provided information. In particular, disclosing the phone number was very effective to attract public engagement which increased likes by 16\%, comments by 14\%, and shares by 33\%; (3) Elucidating the \textit{danger} that victims faced was beneficial to receive more public attention. Specifically, disclosing victims' vulnerability (e.g., children and seniors getting affected), describing the disaster severity (e.g., rising water level), mentioning the affected population size, and showing the possibility of losing communication all contributed to a higher chance of being shared. One exception was the supply shortage which reduced public engagement. A possible reason was that many posts indicating supply shortage asked for supply support instead of rescue; (4) The \textit{subjectivity} when expressing help-seeking requests could arouse public attention. However, three types of negative \textit{emotions} (anger, anxiety and sadness) surprisingly did not promote public engagement, which contradicted prior findings focusing on the health crisis~\cite{luo2020triggers}. 

\subsubsection{Non-linguistic Strategies}

Structuring the posts with a normalized syntax (\emph{structure}, IRR=1.06, p<0.001), adopting figures or videos (\emph{multi\_media}, IRR=1.39, p<0.001) and adding disaster-specific hashtags (\emph{hashtag}, IRR=1.39, p<0.001) all enhanced the engagement and spread of help-seeking posts, which indicated these strategies' effectiveness in attracting public attention. On the contrary, the effort of mentioning others (\emph{social\_connection}, IRR=0.93, p<0.001) failed to promote the public engagement.

\subsubsection{Control Variables}

Apart from how the linguistic and non-linguistic strategies, the intrinsic needs and contexts of help-seeking naturally influenced public engagement. Compared to the posts seeking supply, people were engaged more in posts seeking rescue. Users with stronger social ties received more attention when looking for help, when having one more follower brought a 157\% increase of the total engagement index. \textit{Information density} (IRR=0.75, p<0.001) negatively predicted social media engagement. As such, the findings on contextual features largely distinguished seeking help via broadcasting from seeking help via specific targeting (i.e., asking for help from authoritative sources), when the latter was barely impacted by the social ties of the help seekers and the level of busyness~\cite{li2021hello}.

\subsection{RQ3: Community's Work to Prevent Help-Seeking Overwhelm}\label{communityWork}

In this section, we describe how users in the community collaboratively made an effort to prevent help-seeking posts from being overwhelmed. Such community's work included \textit{norm development}, \textit{norm broadcast}, \textit{norm enforcement}, \textit{raising attention for less-noticed posts} and \textit{cross-community support}. 

\subsubsection{Norm Development}\label{normDevelopment}

When a specific post raised the concern on the help-seeking overwhelm, community members spontaneously initiated discussions in the comment, aiming to reach a consensus on norms that could promote the visibility of individuals' help-seeking posts. They raised negative consequences when help-seeking requests were not properly conveyed, put forward possible solutions, and complemented each other to make norms comprehensive. Here is an example:

\begin{quote}
    \textit{\textbf{Poster:} \#Henan Rainstorm Mutual Aid\# The help-seeking information is too repetitive and mixed. It increases the workload to the staff who gather rescue information, and makes the front line hard to rescue... Please forward the original post instead of copying the content to your microblog. I know everyone is worried, but don't spoil things with good intentions.\\
    \textbf{U1:} Agree. Only when you or your family need help, or someone clearly asks you to post it, you can post the original help-seeking information in the (Henan Rainstorm mutual help) Super Topic. \\
    \textbf{U2:} And after being rescued, please delete the microblog and leave the help-seeking channel for others.}
\end{quote}

In this example, the origin poster raised the concern about repetitive help-seeking posts, and initiated the norm to avoid copying and sharing. U1 better clarified the norm under the specific context, and U2 supplemented it with another possible solution that argued deleting the microblog after being rescued. Such processes, under the wisdom of the crowd, helped to generate inclusive norms to maximally prevent help-seeking information from being overwhelmed under the current design of the microblogging platform.

\subsubsection{Norm Broadcast}

We observed that a set of norms were generated and broadcast to enhance the visibility of help-seeking posts as shown in Table \ref{tab:norms}. They involved suggestions for help-seekers (e.g., structure help-seeking information), sharers (e.g., keep the originality when sharing) and other community members (e.g, ban irrelevant comments), and covered both before-posting and after-posting periods. Some norms emerged through the discussion of community members as mentioned in Section \ref{normDevelopment}, and some norms were developed based on the work of digital volunteers. For instance, the example for the norm "\textit{include critical details}" in Table \ref{tab:norms} was concluded when digital volunteers found that some help-seekers did not disclose phone numbers, which might lead to the failure of getting responses. 

\begin{table*}
  \small
  \caption{Norms built by the community to prevent help-seeking posts from being overwhelmed}
  \label{tab:norms}
  \begin{tabular}{p{1cm}p{2.3cm}p{2cm}p{5.8cm}}
    \toprule
    Period & Norm Type & Target & Example \\
    \midrule
    Before Posting & Use the hashtag and super topic to seek help & Help-seekers & \textit{Compatriots in Henan who are affected by the storm and have difficulties can post help-seeking microblogs on Weibo *with the topic included*. Volunteers will collect the information based on the topic.}\\
    & Structure help-seeking information & Help-seekers & \textit{...You can use the \#Henan Rainstorm Mutual Aid\# topic to request help. It is recommended to structure your posts with [address:], [contact information:], and [description of your difficulty:].}\\
    & Include text description & Help-seekers & \textit{Please include the text version of your help-seeking request, instead of only in the image, so that people in network-affected areas can load and see.}\\
    & Include critical details & Help-seekers & \textit{My friends, when you request for help and supplies, please include your phone number. Otherwise, we can not report it to the rescue team.}\\
    & Keep the originality when sharing & Sharers & \textit{When you want to spread the help-seeking information, please forward the original microblog, instead of copying and posting by yourself. Otherwise it will lead to the repetitive rescue. Thanks for your collaboration.}\\
   
    \hline
    After Posting & Delete posts after getting rescued  & Help-seekers and sharers & \textit{I hope that those who have been rescued can report safety and delete their microblogs to leave more resources for people who are still waiting for help. }\\
    
    \hline
    Anytime & Ban irrelevant comments (e.g., insult and conspiracy talk) & All community members & \textit{Never be a troll. You are occupying the public space on this topic and influencing others who need help!}\\
    & Raise attention on help-seeking posts in less represented areas  & All community members & \textit{Now the focus here is still Zhengzhou. Actually the situation in Zhengzhou is stable now, while many people in Xinxiang need help. Please do not ignore their request for help!}\\
    \bottomrule
  \end{tabular}
\end{table*}

\subsubsection{Norm Enforcement}

Community members, including some digital volunteers, collaboratively worked to enforce norms through commenting under non-standard help-seeking posts, and thus promoted their chance of being responded instead of getting neglected. Here is an example:

\begin{quote}
    \textit{\textbf{Poster:} \#Henan Rainstorm Mutual Aid\# Need urgent help!\\
    (detailed address) in Sizhuangding Village, Xinxiang. Here are several families. A total of 11 people, including one suffering from heart disease! A lady has collapsed!
    \\
    \textbf{U1:} Phone Number please.\\
    \textbf{U2:} Please provide the name and contact telephone number!\\
    \textbf{Poster:} Contact: (Phone Number)\\
    \textbf{U3 (Volunteer):} Reported to the rescue team by online volunteers.}
\end{quote}

In this example, when the poster requested urgent help without providing contact information, community members repeatedly reminded the poster till the digital volunteers were able to report the complete help-seeking information. Another typical example showed how users tried to enforce the norm "\textit{delete posts after getting rescued}" (though their effort failed in this case as we collected this post after the disaster):

\begin{quote}
    \textit{\textbf{Poster:} \#Henan Rainstorm Mutual Aid\# The whole Qimen Village in Xinzhen Town, Xun County is surrounded by water. The embankment is about to break down. Now villagers have to perform self-rescue. There is no rescue team, and we can not reach them. Please help us!!! There are 8000 people in our village, with many old people and children.\\
    (Name), (Phone Number), 8:36\\
    \textbf{U1:} Troops have been there. They arrived at about 1:00 this morning.\\
    \textbf{U2:} Rescue workers are already there. If you are safe now, please delete the post.}
\end{quote}

\subsubsection{Raising Attention for Less-Noticed Posts}\label{raisingAttention}

Community members voluntarily left supportive comments under unsolved or unanswered help-seeking posts to bring public attention to them. The most common example was "\textit{up!}", when community members simply aimed to keep the activity level of specific posts which were about to be submerged. More elaborate strategies were also perceived. For instance, some community members copied and pasted less-noticed posts in the comment of some popular posts to attract attention. In another case, community members collaboratively leveraged the "@" interface to mention rescue teams and digital volunteers that they knew in the comment of less-noticed posts to increase the chance of being noticed.

\subsubsection{Cross-Community Support}

A surprising type of community's work to prevent help-seeking overwhelm is the cross-community support. Specifically, when some users retweeted or commented on the help-seeking posts, they added hashtags or super topics of other popular communities, typically regarding celebrities with a large number of followers, to attract attention from these communities. Here is an example:

\begin{quote}
    \textit{\textbf{Poster (Sharer):} \# Super Topic of (A Celebrity's Name)\# My friends following the (celebrity's name), please help me forward it! In a hurry, but still not responded. // \#Henan Rainstorm Mutual Aid\# SOS! (A list of village names and addresses.) All villages are badly flooded. No food, no power, no signal. The flood dikes are all broken, and people are staying on the roof. We need rescue troops!!! Civilian rescue teams are not suggested to come here. It is too dangerous.\\
    \textbf{A list of replies:} Forwarded!
    }
\end{quote}

In this example, the sharer added a super topic with about 10 million followers and aimed to get support from that community. The post was then widely spread. However, whether this strategy led to another cause of imbalanced attention was still unclear.

\section{DISCUSSION}

In this work, we comprehensively investigated the challenge of help-seeking overwhelm during a natural disaster, and uncovered strategies developed by individuals and communities for the overwhelm prevention. This section discusses how our findings expand the understanding of help-seeking during natural disasters and shed light on implications to support more effective and efficient help-seeking behaviors. We first reflect on the visibility challenges of help-seeking requests under massive and miscellaneous crisis communication in Section \ref{discussion-I}, proposing design implications to hinder external and internal overwhelm at the system level. Then, in Section \ref{discussion-II}, we highlight the essential role of individuals' help-seeking patterns, which enlightens practices to facilitate efficacious help-seeking requests and warns of the disastrous consequences of exaggeration. In Section \ref{discussion-III}, we think over how to better leverage the power of online communities to afford the resilience of help-seeking overwhelm.

\subsection{Under Massive and Miscellaneous Posts: Reflecting on the Visibility of Help-Seeking Requests}\label{discussion-I}

Prior work has emphasized the significance of help-seeking posts during natural disasters to efficiently acquire victims' urgent needs and better organize the rescue~\cite{cheng2020explaining,li2019using, keim2011emergent,nishikawa2018time, song2019toward}. Nevertheless, under the ``information explosion'' on social media after disasters, the visibility of help-seeking posts is not optimistic~\cite{nishikawa2018time}. By showing evidence of both internal and external overwhelm of help-seeking posts, this work not only enriches the understanding of help-seeking requests online, but also raises design implications to create a better environment for help-seeking.

First, this work found that even in posts containing the keyword related to help-seeking (\textit{\#Henan Rainstorm Mutual Aid\#} in this case), information was still \textbf{miscellaneous}, and the pure help-seeking content was not the majority. As shown in Section \ref{ExternalResult}, \textit{help-seeking} posts only accounted for about 40\% in the dataset, which were fewer than posts \textit{sharing attitudes and opinions}. That dovetails with previous work which indicated that most posts with the rescue-related hashtag during disasters were actually NOT for help-seeking~\cite{nishikawa2018time}. Indeed, rich nuances are added to specific topics when people gather on social media during natural disasters~\cite{palen2007citizen}, whose crisis communication needs vary extensively~\cite{qu2009online,qu2011microblogging}. Meanwhile, our findings further warned of the inadequate attention and engagement of \textit{help-seeking} posts compared to other categories like \textit{offering help} or \textit{transmitting critical information}. For instance, the \textit{help-seeking} posts received the fewest likes and comments among the four identified categories. Though other categories of posts could satisfy other information needs during crises, the miscellaneous information largely hindered help-seeking posts from promptly reaching the intended audience. In this regard, assigning subcategories to posts that contain help-seeking-related keywords, whether human-generated through crowdsourced tagging or machine-generated via automatic text classification, would be promising to help online and offline volunteers find those in need more efficiently. Also, it is important for future work to comprehend why \textit{help-seeking} posts fail to get sufficient attention compared to other types of information. As prior work indicated, how people reacted to information during crises was shaped by a complex emotional, perceptive, and cognitive process~\cite{lifang2020effect, berger2012makes,jin2010role}. Understanding information receivers' perceptions that guide their behaviors when processing help-seeking requests information would be beneficial to draw practical implications with more effective and engaging help-seeking strategies.

Second, this work also uncovered how help-seeking on social media might fail under the \textbf{massive} volume of help-seeking posts in a short period. As exhibited in Section \ref{internalResult}, more than 95\% of help-seeking posts shared less than 5\% user engagement (likes, comments and shares), and about 40\% posts received no response. This finding provided the new evidence of \textit{attention inequality} in social media~\cite{zhu2016attention, van2016employing} in the setting of crisis communication, which was even worse compared to the general case~\cite{zhu2016attention}. Though it is natural given the varied social capital of help-seekers, we warn of the catastrophic consequences of such attention inequality. Some popular help-seeking posts are disseminated broadly even after the rescue, while the unsolved help-seeking requests remain neglected. Also, challenges of filtering and retrieving are raised for digital volunteers' self-organized work on collecting and routing information in crisis~\cite{kaufhold2016self,hughes2015social}. Thus, we urge for reflections on how online help-seeking systems can be specially designed to make the overwhelmed posts more visible. For instance, typical ranking and recommendation algorithms might lead to popularity bias (i.e., popular items are ranked highly and recommended frequently) and amplify attention inequality~\cite{abdollahpouri2019popularity, baeza2018bias}. On this note, specialized recommendation systems, giving a higher priority to less noticed help-seeking posts, shall be considered and tested by researchers and designers.

\subsection{Help-Seeking Strategies Matter: Implications and Warnings}\label{discussion-II}

Effectively expressing the help-seeking needs and requests is regarded as an essential approach to attract attention from the public and facilitate the information extraction for digital volunteers~\cite{li2021influence,starbird2011voluntweeters}. Driven by communication theories (e.g., negativity bias theory~\cite{rozin2001negativity} and social support theory~\cite{vaux1988social}), prior work broadly investigated the impact of content and creator factors on the response and dissemination of help-seeking posts~\cite{li2021influence, luo2020triggers,lifang2020effect}. However, when nuanced strategies are developed and adopted by individuals for help-seeking, a deeper understanding of the strategies, beyond the high-level content type and emotion category~\cite{li2021influence,lifang2020effect,li2021hello}, is warranted. This work establishes a comprehensive taxonomy of linguistic and non-linguistic strategies that influences the public engagement of help-seeking posts. We reflect on the effective help-seeking practices and how platforms shall support the articulation of help-seeking needs, and warn of the exaggeration and misrepresentation when help-seeking strategies matter.

This work well demonstrated the ``power of language''~\cite{ng1993power} in the scenario of help-seeking. As shown in Section ~\ref{influencingMechanism}, including sufficient rescue-relevant details (e.g., phone number), disclosing danger (e.g., population vulnerability and potential of losing communication), and using subjective narratives all contributed to a higher chance of getting responded and spread. However, not every victim managed to leverage these strategies to explain their danger and needs, which might aggravate the attention inequality. Recall that community-developed norms urged for ``\textit{including critical details}'' when many victims forgot to attach their phone number in the post. Additional labor was taken by the community or digital volunteers to achieve situational awareness and acquire sufficient rescue-relevant details, which reduced the efficiency of crisis communication. Consequently, future researchers and designers shall think over and investigate how help-seeking-specific interfaces can support valuable information disclosure in the \textit{drafting and posting} stage of help-seeking, supporting efficient information extraction and expediting the rescue plan making.

Under the current design of social media platforms, non-linguistic strategies also had substantial influences on public engagement. For instance, structuring posts to a normalized syntax, similar to the effort of Tweak the Tweet~\cite{starbird2012promoting,starbird2011voluntweeters}, positively predicted the number of likes, comments, and shares. The promotion effects also held for attaching hashtags to signify the help-seeking purpose, and adopting figures and videos to vividly describe the disaster situation. These findings empirically validated the effectiveness of the naturally developed crisis communication principles (e.g., hashtag use) from a quantitative perspective~\cite{olteanu2015expect}. To this end, system-facilitated construction of help-seeking posts, e.g., providing an elaborately-designed template for post structuring, shall be considered by future researchers and practitioners. Meanwhile, we suggest more in-depth investigations on the non-linguistic patterns which are less understood in the existing literature. For example, when video content and forms intrinsically influence public engagement in crises~\cite{he2022more,chen2021factors}, understanding which videos are shared in help-seeking posts and how they influence the public response is a promising research direction.

We finally warn of the disastrous consequences of exaggeration or misrepresentation when specific help-seeking strategies work. For instance, Section \ref{influencingMechanism} indicated that emphasizing situational vulnerability and disaster severity was effective to draw public attention. As such, maliciously misusing such strategies to distract engagement from those in real need would be catastrophic and shall be particularly focused on for future researchers. When misinformation has been a significant topic in crisis communication~\cite{yang2021know, huang2015connected,zarocostas2020fight}, we call for more research attention into how to cope with misinformation in the help-seeking context.

\subsection{Community's Effort for Overwhelm Prevention: Insights and Facilitation}\label{discussion-III}

How online communities facilitate crisis communication is gaining growing attention in HCI and CSCW~\cite{xu2018crisis, qu2009online,qu2011microblogging,huang2015connected, starbird2013delivering}. In the specific setting of help-seeking, we revealed that community members also collectively developed countermeasures and collaboratively worked to prevent the help-seeking posts from being overwhelmed. This section discusses (1) how the community's effort for overwhelm prevention provides insights into the design for effective help-seeking, and (2) how the design can be in turn improved to facilitate the community's collaborative work.

``The wisdom of the crowd'' plays a crucial role in crisis communication such as debunking misinformation and conspiracy~\cite{tanaka2013toward, yang2021know, he2021beyond, arif2017closer}. Our findings contribute to this line of work by showing how the public collaboratively came up with strategies for overwhelm prevention and put them into practice. For example, through a process of discussion and negotiation, community members thought out a set of norms such as \textit{keeping the originality when sharing} and \textit{deleting posts after getting rescued}, and further disseminated and enforced the norms under the current design of Weibo. These crowd-developed norms were surprisingly comprehensive, which covered different periods (before and after posting) and different targets (help-seekers, sharers, or all community members). Generally, these norms reflected the essential needs of help-seekers and digital volunteers to cope with the information overwhelm, and provided significant insights into the design of help-seeking communities. When community norms are increasingly valued to help the community achieve its goals~\cite{kiesler2012regulating, chancellor2018norms}, future designers and researchers shall actively reflect on the insights of crowd-developed norms and propose corresponding interfaces to provide system-level support. For instance, automatic reminders for help-seeking post deletion after successful rescue could potentially complement the current user-initiated norm enforcement and make the overwhelmed posts surface.

In return, we underscore the necessity of better situating the community's collaborative work in the help-seeking community's original function, instead of letting them evolve as two processes interfering with each other. For instance, as indicated in Section \ref{normDevelopment}, current collective brainstorming on norms regarding help-seeking typically tagged the posts with help-seeking-related keywords to attract community members for discussion. However, when such discussion was mixed with help-seeking posts, users' attention was dispersed, i.e., users needed to pay additional effort to distinguish and trace either norm discussion or help-seeking requests. Another typical example was that with the community's effort to raise attention for the ignored help-seeking requests (see Section \ref{raisingAttention}), a large volume of repetitive and less meaningful posts and replies were generated (e.g., "\textit{up!}"), which also complicated the fulfillment of communities' original purpose of help-seeking. In this regard, future designers are recommended to bridge the gap between \textit{the community's natural countermeasures to prevent overwhelm} and \textit{insufficient support to facilitate effective cooperation}. For example, a separate zone for norm discussion is warranted to avoid entanglements between pure help-seeking requests and norm developing effort.

\subsection{Limitations and Future Work}

This work performed a mixed-methods approach to investigate the overwhelm situation and individuals' and communities' corresponding strategies of help-seeking during a natural disaster, shedding light on considerations for efficient and effective help-seeking behaviors online. However, this work suffered from the following limitations: (1) as we focused on help-seeking posts after one recent natural disaster in a Chinese social media platform, the findings may not be generalized to other platforms with different designs or other disasters with different situational conditions; (2) the community-developed norm for \textit{deleting posts after getting rescued} reduced a portion of data, which might introduce a bias into the study and slightly influence the results; (3) due to the difficulty to quantitatively measure the online communities' strategies, we did not investigate their effects on the overwhelm prevention. 

When the help-seeking overwhelm is a critical issue in crisis communication yet related work is still limited, we call for more in-depth investigations into this topic. Future work shall compare help-seeking behaviors across different platforms to understand how the platform design may influence the help-seeking process, and raise proof-of-concept interfaces to thoroughly evaluate the design implications. Also, a qualitative study on how users perceive different help-seeking strategies and make corresponding responses would be beneficial to draw valuable guidelines for effective help-seeking.

\section{CONCLUSION}

This work adopts a mixed-methods approach to investigate the situation of help-seeking overwhelm on a popular Chinese social media platform Weibo during 2021 Henan Floods, and how individuals and communities develop countermeasures to cope with it. We find that help-seeking posts face critical challenges to get adequate public engagement. They might not only be overwhelmed by massive non-help-seeking posts that also include the help-seeking-related keyword, but also be submerged under the attention inequality of help-seeking posts (less than 5\% help-seeking posts attracting more than 95\% likes, comments and shares). We extract a comprehensive taxonomy of linguistic and non-linguistic strategies of help-seekers to promote public attention, and explore their influences using negative binomial regression models. The results indicate that including contact information, describing danger and vulnerability, adopting subjective narratives, structuring the post to a normalized syntax, and adopting hashtags and multi-media all help to enhance the engagement of help-seeking requests, while expressing negative emotions and mentioning others have no promotion. Finally, a qualitative content analysis uncovers the community's spontaneous effort to prevent the overwhelm, which involves both collective wisdom (e.g., norm development through discussion) and collaborative work (e.g., norm broadcast and enforcement). Based on the findings, we propose design implications that support effective help-seeking during natural disasters.

\section{Acknowledgments}
The research was supported in part by RGC RIF grant R6021-20, and RGC GRF grants under the contracts 16209120 and 16200221.

\bibliographystyle{ACM-Reference-Format}
\bibliography{sample-base}

\end{document}